\documentclass[12pt]{article}
\usepackage{titling}
\title{\textbf{Axially symmetric collapses  in the 2-D Benjamin-Ono equation}}
\author{\textbf{Joseph O. Oloo$^{1,2}$} \\ \& \\ \textbf{Victor I. Shrira$^1$}\\[1.0em] \large $^1$ School of Computing Engineering and Built Environment, \\EH10 5DT Edinburgh, United Kingdom\\$^2$ School of Computing and Mathematics, Keele University \\ST5 5BG Staffordshire, United Kingdom. \\[0.5em]\textbf{j.oloo@napier.ac.uk}\\[0.5em]\textbf{v.i.shrira@keele.ac.uk}}

\usepackage{latexsym}

\usepackage{amsmath,amssymb}
\usepackage{fullpage}
\usepackage{hyperref}
\usepackage{graphicx}
\usepackage{setspace}
\usepackage{fullpage}
\usepackage{caption}
\usepackage{subcaption}
\usepackage{animate}
\usepackage{tabularx}
\usepackage[export]{adjustbox}
\newcommand{\be}{\begin{equation}}
	\newcommand{\ee}{\end{equation}}

\begin{document}
	\begin{titlingpage}
		\maketitle
		\vspace{-2em}
		\begin{abstract}
We study the nonlinear dynamics of localized perturbations within the framework of the essentially two-dimensional generalization of the Benjamin-Ono equation (2D-BO) derived asymptotically from the Navier-Stokes equation.  By simulating the 2D-BO equation with the pseudospectral method, we confirm that the localized initial perturbations exceeding a certain threshold collapse, forming a point singularity.  Although the 2D-BO equation does not possess axial symmetry, we show that in the vicinity of the collapse singularity, the solution becomes axially-symmetric, whatever its initial shape.  We find that perturbations collapse  in a self-similar manner, with the  perturbation amplitude exploding as $ (\check \tau)^{-\lambda}$ and its transverse  scale shrinking as  $ (\check \tau)^{\lambda}$, where $\check \tau$ is the time to the moment of singularity. We derive a  family of self-similar solutions describing axially symmetric collapses. The value of the free parameter $ {\lambda}$ in the self-similar solution is specified by fitting it to the numerical simulation of the initial problem of the evolution of an initially localized perturbation.  Remarkably, for the examples we examined the value of the parameter proved to be almost universal: $ {\lambda} \approx 0.9$; its dependence  on the initial conditions is indiscernible. In the vicinity of the singularity, the dynamics becomes one-dimensional, thus, the derived reduction of the 2D-BO equation provides an effectively  one-dimensional model of collapse. 	
 		 \end{abstract}
	\end{titlingpage}
	\pagenumbering{gobble}
	\pagenumbering{arabic}
	
	\section{Introduction}
We study the nonlinear evolution of localized perturbations within the framework of the essentially
two-dimensional generalization of the `Benjamin-Ono equation' (2D-BO).

A family of nonlinear evolution equations describing   essentially three-dimensional (3-D) long-wave perturbations in a 	free-surface boundary layer with and without density stratification and various explicit account of viscous damping was derived and examined \cite{shrira1989subsurface}, \cite{voronovich1998two}, \cite{oloo_shrira_2023}). Of particular importance is the simplest model derived  for homogeneous deep fluid under assumption of unidirectional basic flow (Shrira 1989) governed  by the (2+1)-D Benjamin-Ono (2D-BO) equation
	\begin{equation}\label{IntroEvolutionEquation1}
		A_{\tau}+ AA_{x}-\hat{G}[A_{x}]		=0,
	\end{equation}
where $A(x,y,\tau)$ is the amplitude of the `vorticity mode' 	dependent on the streamwise and spanwise variables, $x,\,y$ and slow time $\tau$; the nonlocal operator $\hat{G}$ describing the  mode dispersion is
	\begin{equation}\label{Intro-eqGintegralOperator1}
		\hat{G}[\varphi(\boldsymbol{r})]=\frac{1}{4\pi^{2}}
		\int_{-\infty}^{+\infty}\int_{-\infty}^{+\infty}|\boldsymbol{k}|
		\varphi(\boldsymbol{r}_{1})e^{-i\boldsymbol{k}(\boldsymbol{r}-\boldsymbol{r}_{1})}
		\mathrm{d}\boldsymbol{k}\mathrm{d}\boldsymbol{r}_{1}.
	\end{equation}
Operator $ \hat{G}[\varphi(\boldsymbol{r})]$ describes
the essentially  two-dimensional Benjamin-Ono type
$|\boldsymbol{k} |$ dispersion.  Here  $\boldsymbol{k}=(k_x, k_y) $ is the wave vector, $\boldsymbol{r}= (x,y)\,, $ and $\varphi$ is an arbitrary scalar function. We  also use an equivalent alternative representation of $\hat{G}$ in terms of the hypersingular Cauchy-Hadamard 	integral
	\begin{equation}\label{Intro-eqGintegralOperatorCauchy}
		\hat{G}[\varphi(\boldsymbol{r})]= \frac{1}{2\pi}\int_{-\infty}^{\infty} \int_{-\infty}^{\infty}	 \frac{\varphi(x^{\prime},y^{\prime},\tau)\,dx^{\prime}\,dy^{\prime}}{[(x-x^{\prime})^{2}+(y-y^{\prime})^{2}]^{3/2}}.
	\end{equation}
The improper integral  (\ref{Intro-eqGintegralOperatorCauchy}) 	is understood as the Hadamard finite-part integral.

The  two-dimensional equation (\ref{IntroEvolutionEquation1}) describes evolution of three-dimensional perturbations in the following sense: in the asymptotic procedure of derivation of \ref{IntroEvolutionEquation1}  the perturbation dependence on vertical coordinate $z$  to leading order splits off; is described by explicit solution of linear boundary value problem, while the  dependence on horizontal coordinates $x,y$  and time is governed by  the 2D-BO equation on amplitude $A$.   $A$ is a normalized amplitude of the perturbation  streamwise velocity component. The  detailed derivation of  (\ref{IntroEvolutionEquation1}) and its generalizations can be found in ( \cite{voronovich1998two}, \cite{oloo_shrira_2023}). We stress that in contrast to the 	Kadomtsev-Petviashvilli (KP) type models, the 2D-BO 	equation does not assume any smallness of transverse wavenumber ($|k_{y}|$), it is an generalization of the  Benjamin-Ono equation similar to the KdV extension in \cite{zakharov1974three}.  The importance of studying the  2D-BO equation is in the fact that it is the limit of a broad class of geophysically relevant models (e.g. \cite{oloo_shrira_2023},  \cite{oloo2020boundary}, \cite{voronovich1998two}).

In the limit of strictly planar perturbations  the original 2D-BO equation 	 \eqref{IntroEvolutionEquation1}    reduces to the classical one-dimensional Benjamin-Ono equation (BO), which  is one of the  universal integrable  nonlinear long-wave evolution equation emerging in various physical contexts. The  BO equation was originally derived 	for  long internal waves in deep stratified 	fluid
	\cite{benjamin1967internal}, \cite{davis1967solitary} and \cite{ono1975algebraic}. The  BO equation  is integrable in the sense, 	that it conserves an infinite  set of integrals of motion; it possesses both multi-soliton and multi-periodic wave 	solution and  it also exhibits other properties typical of
integrable systems (e.g., 	\cite{case1978n},\cite{case1979properties}, 	\cite{ablowitz1981solitons},
\cite{matsuno1984bilinear}).  In  the one-dimensional 	setting the solitons are very robust.

In contrast to the  thoroughly examined  BO equation, the 2D-BO 	equation is not so well studied.
The  obvious vast class of exact solutions of the 2D-BO 	equation  (and its KP reduction), which comprises all steady and evolving oblique plane wave solutions of the one-dimensional Benjamin-Ono (BO) equation, 	 is unstable with respect to long transverse perturbations \cite{pelinovsky1994self}. 	A more detailed analysis of this transverse instability was carried out in 	 \cite{gaidashev2004transverse}. The 2D-BO 	equation is not axially symmetric, however, counterintuitively, 	 it admits  steady axially symmetric solitary wave 	solutions \cite{abramyan1992multidimensional}.      These steady solutions were initially thought to be prototypes of 3D coherent soliton-like structures, but later proved to be also unstable. Their role in the dynamics of the 2D-BO solutions has not been clarified yet. Here, we revisit these  solutions and will shed light onto this gap.
	
The crucial advance in the understanding of dynamics within the 2D-BO equation is due to \cite{d1995two}. In \cite{d1995two} Dyachenko \& Kuznetsov (1995) showed that the 2D-BO equation describes collapses: i.e. the amplitude of a collapsing perturbation becomes infinite in finite time, while its width shrinks to zero. It has also been 	 suggested that the   equation possesses self-similar solutions which describe emergence  of a singularity, however a full analysis of self-similar solutions has not been carried out.    An explicit description of the collapse evolving as a result of the transverse instability of a plane solitary wave was derived employing Whitham's adiabatic approach  within the framework of the KP reduction of the 2D-BO equation in \cite{pelinovsky1995collapse}.
	
Here, by means of  numerical simulations of  the 2D-BO equation we confirm theoretical prediction that all initially localized perturbations exceeding a certain threshold  collapse. We also find that the collapsing perturbations tend to become axially symmetric in the vicinity of the singularity, whatever the shape of the initial perturbation. Why do the arbitrarily shaped initial localized perturbations of sufficient amplitude tend to become axially symmetric within the evolution equation lacking the axial symmetry? At present it is a mystery. To check the idea that  this happens due to the self-similarity of the evolution of wide class of initial conditions we in \S\ref{axial-symmetry} examine self-similar solutions  in the vicinity of the singularity. The self-similar solutions are not necessarily axially symmetric, but we find a class of axially symmetric solutions describing  the vicinity of the singularity. Moreover, we demonstrate that a particular family of this class of axially symmetric solutions emerges as the asymptotics of a wide class of localized initial conditions. We also reveal links between the 2D-BO collapses and the axially-symmetric steady solutions found in  	 \cite{abramyan1992multidimensional}: for the found class of self-similar solutions the instantaneous shape of a collapsing pulse is governed by the \cite{abramyan1992multidimensional} solutions.

The work is organized as follows. A brief mathematical formulation  in  \S\ref{The-Model}, is followed  by
a discussion of numerical simulations of typical scenarios of collapses in  \S\ref{Collapses in 2dBO}. The main point is in demonstrating the overlooked tendency of collapsing pulses towards axially-symmetric self-similar solutions. In section \S\ref{Self-similarity} we study self-similar solutions of the  of the 2D-BO equation, in particular, in \S\ref{axial-symmetry}  we examine a class of axially-symmetric self-similar
solutions in the vicinity of the singularity. To better analyze these solutions we also
transform the two-dimensional Benjamin-Ono dispersion operator from the Cartesian into polar form;
the  details of the algebra are given in the Appendix \ref{G1-operator-transformation}. We find that for the derived self-similar solutions the amplitude time dependence $\tilde\tau^{-\lambda}$ splits-off, while  the instantaneous shape of axially-symmetric collapsing pulses is governed by the  one-dimensional integro-differential equation which is the axially symmetric reduction of the equation numerically examined in \cite{abramyan1992multidimensional}. In \S\ref{Quantative Comparisons} we compare the self-similar solutions with the numerical simulations of a number of initial conditions. We find excellent agreement. Remarkably, to leading order the asymptotic behaviour near the singularity does not  depend on the initial conditions.  In the \emph{Concluding Remarks} (section \ref{Main-Conclusions}) we summarize the results and discuss the open questions.

	\section{Problem Formulation}\label{The-Model}
The underpinning physical model is rooted in the reality of the ubiquitous free surface flows in nature [e.g.\cite{soloviev2013near}]. We consider the evolution of three-dimensional localized finite-amplitude perturbations of a steady unidirectional boundary layer shear flow $\boldsymbol{U}$ adjacent to an 	infinite free-surface boundary assumed to be flat.
	\begin{figure}[h!]
		\begin{center}
				\includegraphics[width=0.90\textwidth]{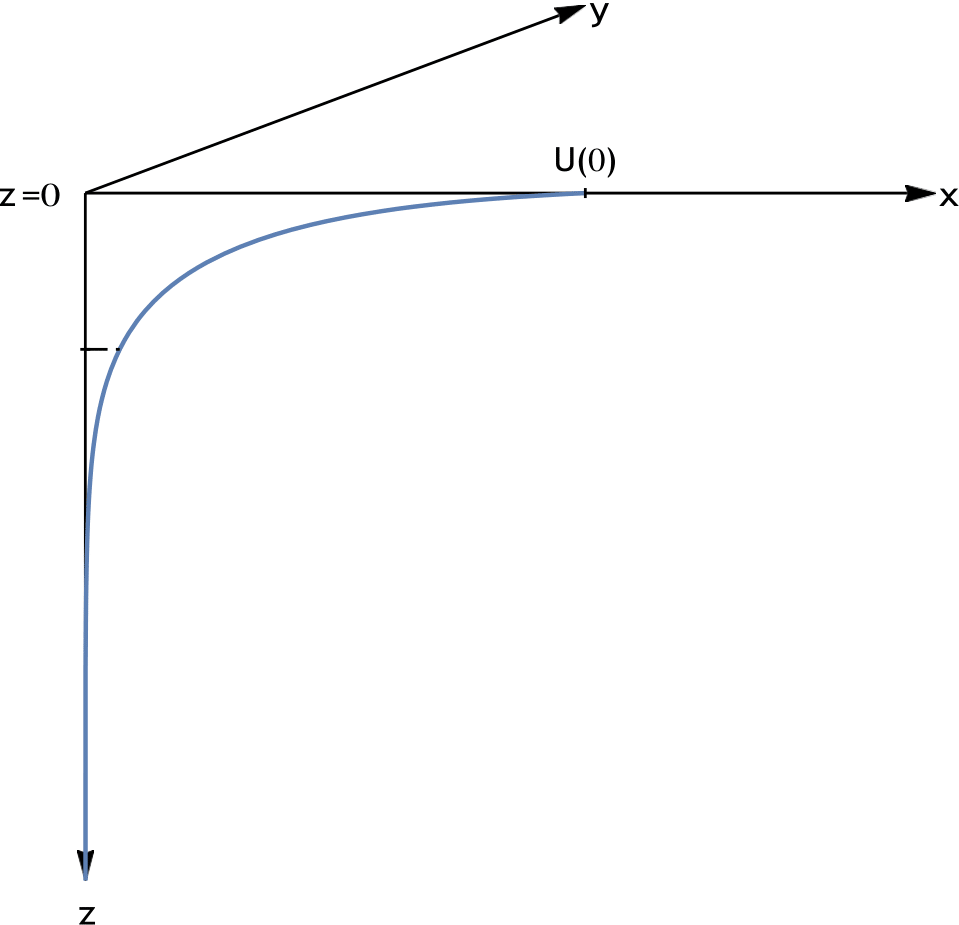}
		\end{center}
		\caption{Sketch of geometry of a typical free-surface boundary
			layer profile. There are no assumptions regarding the profile. Usually, free-surface boundary
			layer has the maximum of velocity
			at the surface $U_{max}=U(0)$ but this point is immaterial for the study.   }
		\label{Geometry-homogeneous}
	\end{figure}
In the Cartesian frame with the origin at the upper boundary with $z,\, 0\leq z\leq \infty$, directed
downwards with $x$ and $y$ directed streamwise and spanwise, respectively (see Fig \ref{Geometry-homogeneous}) we assume the unperturbed unidirectional boundary layer  	$U(z)$ to be  localized in a  layer of characteristic thickness $\varepsilon$  small compared to characteristic wavelength of the perturbations. Assuming weak nonlinearity of the perturbations which we characterize by the same small parameter $\varepsilon$. We link  $\varepsilon$ to the Reynolds number $Re$ assuming it to be  $O(Re^{-2})$, where $Re (Re \gg 1)$. Under  these assumptions from the Navier-Stokes equations for incompressible fluid follows the nonlinear 2D-BO evolution equation \ref{IntroEvolutionEquation1}, derived by means of matched asymptotic expansion \cite{voronovich1998two, oloo_shrira_2023}.

It is known (see \cite{d1995two}) that the 2D-BO equation \eqref{IntroEvolutionEquation1} can be also cast into the equivalent Hamiltonian form
	\begin{equation}\label{B-eqnHamiltonian2}
		A_{\tau}=\partial_{x}\left[\hat{G}[A]-\frac{1}{2}A^2\right]=\partial_{x}\frac{\delta
			H}{\delta A}, \quad H=\frac{1}{2}I_{1}-\frac{1}{6}I_{2},
	\end{equation}
where the Hamiltonian $H$ is the sum of two constituents $I_1$
	and $I_2$, describing, respectively, dispersion due to the perturbation velocity dependence on wavenumber (the `2D Benjamin-Ono dispersion'), and nonlinearity
	\begin{equation}\label{BO-eqn-Hamiltonian}
		H=\frac{1}{2}I_{1}-\frac{1}{6}I_{2}, \quad
		I_{1}=\int A\hat{G}[A]\, \textbf{dr},\qquad  I_{2}=\int
		A^{3}\, \textbf{dr},\quad \textbf{dr}\equiv dxdy.
	\end{equation}
It is also known (see \cite{d1995two}) that besides the Hamiltonian, the 2D-BO equation (\ref{IntroEvolutionEquation1})  conserves three other integrals: the streamwise and spanwise components of the 	`momentum' $\textbf{P}(P_{x}, P_{y})$  and the  mass flux $M$,
	\begin{equation}\label{integrals}
		P_{x}=\frac{1}{2}\int\int A^{2}\,dxdy,\quad
		P_{y}=\frac{1}{2}\int\int A\phi_{y}\,dxdy,\quad
		(\phi_{x}\equiv A),\quad M=\int\int A\,dxdy.
	\end{equation}
The negativity of the Hamiltonian for a chosen initial perturbation means that that nonlinearity prevails over dispersion in the course of its evolution ending in collapse. Thus,  $H <0$, is the sufficient criterion of collapse \cite{zakharov2012solitons}. Note that for the axially symmetric solitary wave solutions found in 	 (\cite{abramyan1992multidimensional}) the Hamiltonian is 	always negative. In the next section we examine emergence of collapses by means of direct numerical simulations  of the 2D-BO equation.
		
\section{Simulation of collapses in the 2D-BO equation: Emergence of the axial symmetry}\label{Collapses in 2dBO}

\subsection{The Method}
In this section we examine collapses of 	initially localized perturbations (  `lumps') by simulating their
development within the framework of the 2D-BO equation 	 \eqref{IntroEvolutionEquation1}. For certainty we mostly confine our attention  to consideration of the evolution of the  Gaussian pulses stretched in the longitudinal and lateral directions.

\subsection{Numerical simulation}\label{Numerical-solution-2dBO}
To simulate numerically the  evolution  equation
\eqref{IntroEvolutionEquation1} for localized initial
perturbations we use the  pseudo-spectral method (e.g.
\cite{orszag1969numerical} 
and \cite{kopriva2009implementing}).
The employed pseudo-spectral  method with periodic boundary
conditions uses efficient
fast Fourier transform (FFT) routines for handling
dependencies on $x$ and $y$, while for the time evolution
the classic fourth order Runge-Kutta method is employed.

In our context it was found to be optimal to   use a
rectangular box of length ${512\pi}$ and width ${128\pi}$.
This choice provides sufficient domain for the spatial
decay of the localized perturbations we were simulating
and also to allow the perturbation sufficient time to move
in the streamwise direction during the evolution.

It is convenient to  present our evolution equation in the conservation law form,

\begin{equation*}
	A_{\tau}+F_{x}=0,  \qquad F=-\hat{G}[A]+\frac{1}{2}A^{2}.
\end{equation*}

The
integral operator $\hat{G}[A]$ is dealt with in the
Fourier space, while nonlinear terms are considered in the
physical space on collocation points with the `two-third

de-aliasing rule' (e.g.\cite{orszag1969numerical}). The
accuracy of the simulations was controlled by ensuring that
the integrals of motion (\ref{integrals}) remain constant
with the error not exceeding $10^{-4}$.

We first examine a few simple  initial
distributions choosing  the Gaussian ($A_G(x,y)$) pulses,
\begin{equation}\label{B-InitialConditions}
	A_G(x,y)=ae^{-\left(\frac{x^{2}}{2\sigma_{x}^{2}}+\frac{y^{2}}{2\sigma_{y}^{2}}\right)},
\end{equation}
These initial configurations are fully characterized by
just  three parameters: amplitude $a$ and characteristic
half-widths $\sigma_{x}$ and $\sigma_{y}$, which we
refer to just as the `widths'.

It is easy to see that  the Hamiltonian $H$ given by
\eqref{B-eqnHamiltonian2} and its constituent integrals,
$I_{1}$, $I_{2}$, $I_{3}$,  can be expressed in terms of
the perturbation initial amplitude $a$ and perturbation
widths $\sigma_{x}$ and $\sigma_{y}$. To elucidate the role of the lateral/logitudional stretching of the initial pulse we also introduce the
`aspect ratio',  $\alpha={\sigma_{y}}/{\sigma_{x}}$ , as a
measure of this  asymmetry.
We consider evolution of   Gaussian pulses of various amplitudes stretched in the longitudinal or lateral directions, that is for $\alpha</>1$.

\subsection{Evolution scenarios}
 Our simulations of the evolution of localized initial perturbations, which we don't report in detail here, suggest that
the phase space of the evolution equation of
\eqref{IntroEvolutionEquation1} is organized very simply:
all perturbations with amplitudes exceeding the threshold specified by the condition $H=0$ collapse, while the smaller perturbations decay. Thus, collapses  and and the unperturbed basic state are the attractors of the system.  There are no nontrivial steady states, limit circles or chaotic attractors.

\begin{figure}
	\centering
	\includegraphics[width=1.0\textwidth]
	{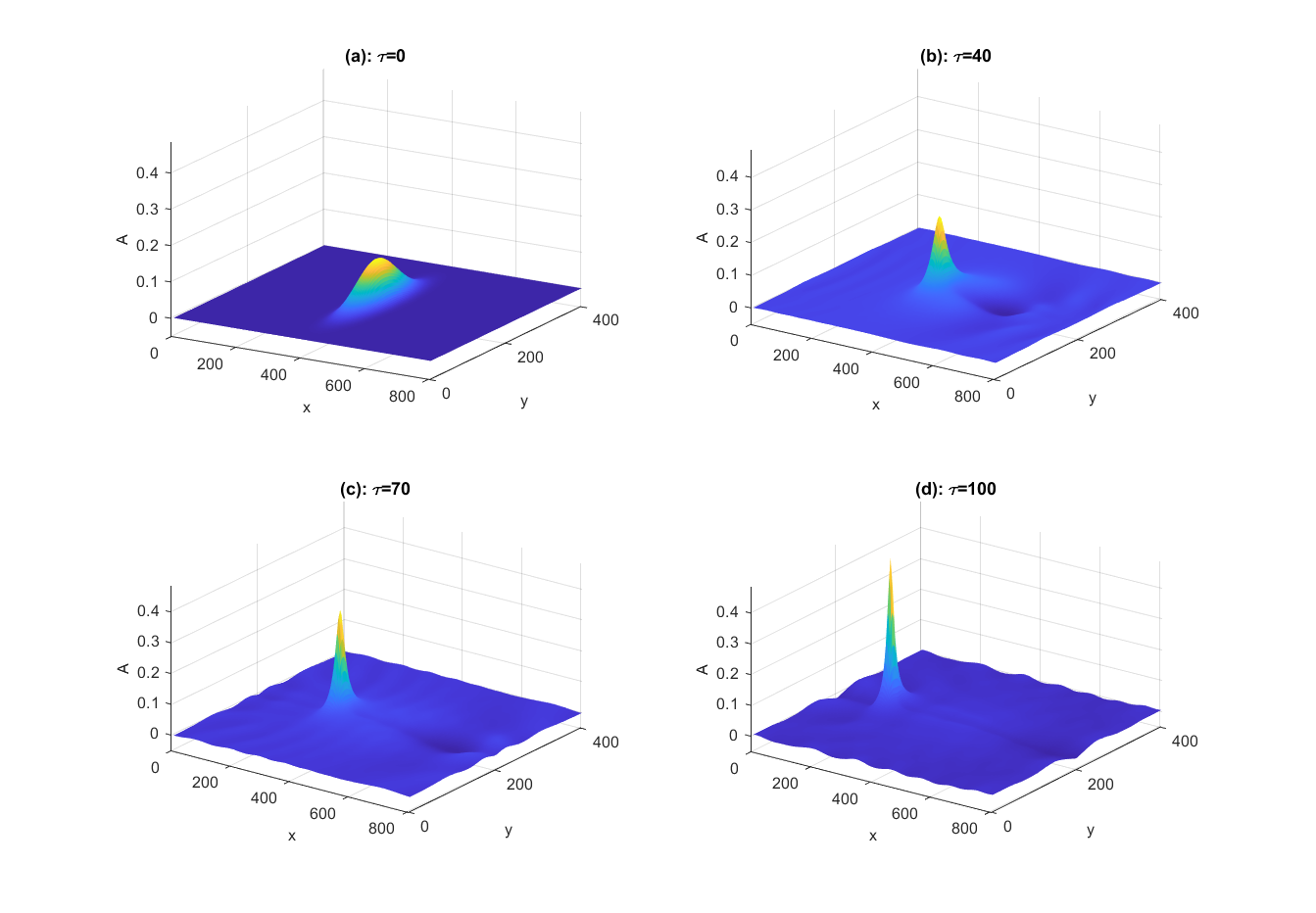}
	\caption{ Snapshots showing the evolution of the amplitude $A(x,y,\tau)$
		of a collapsing initially laterally stretched ($\alpha=2$) Gaussian pulse  taken at four-different moments:  $\tau=0, \, 40, \,70, \,100$. The parameters of the initial Gaussian pulse prescribed by (7) ($a=0.3159,
		\sigma_{x}=25,          \sigma_{y}=50$).
	}
	\label{Collapse-Evolution-BO}
\end{figure}

We  illustrate a typical collapse scenario by providing successive 3-D snapshots of amplitude evolution in figure  (\ref{Collapse-Evolution-BO}) which shows how an initially Gaussian pulse laterally stretched with aspect ratio two  focuses and blows up.

\begin{figure}
	\centering
	\includegraphics[width=1.0\textwidth]
	{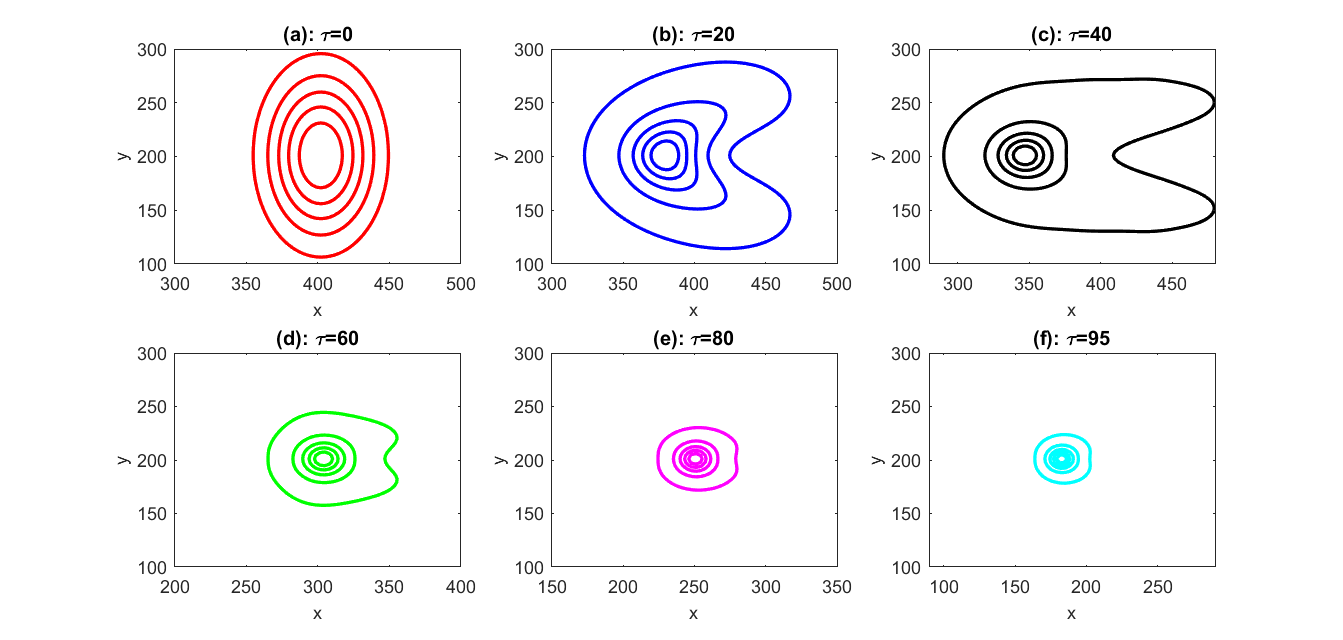}
	\caption{Evolution a collapsing initially
		laterally stretched Gaussian pulse  ($\alpha=2$)  taken at
		six-different moments:  cross-sections. $\sigma_{x}=25,\,\sigma_{y}=50,\,a=0.1353$
	}
	\label{Cross-section_hairpin_vortices}
\end{figure}

\begin{figure}[h!]
	\centering
	\includegraphics[width=1.0\textwidth]
	{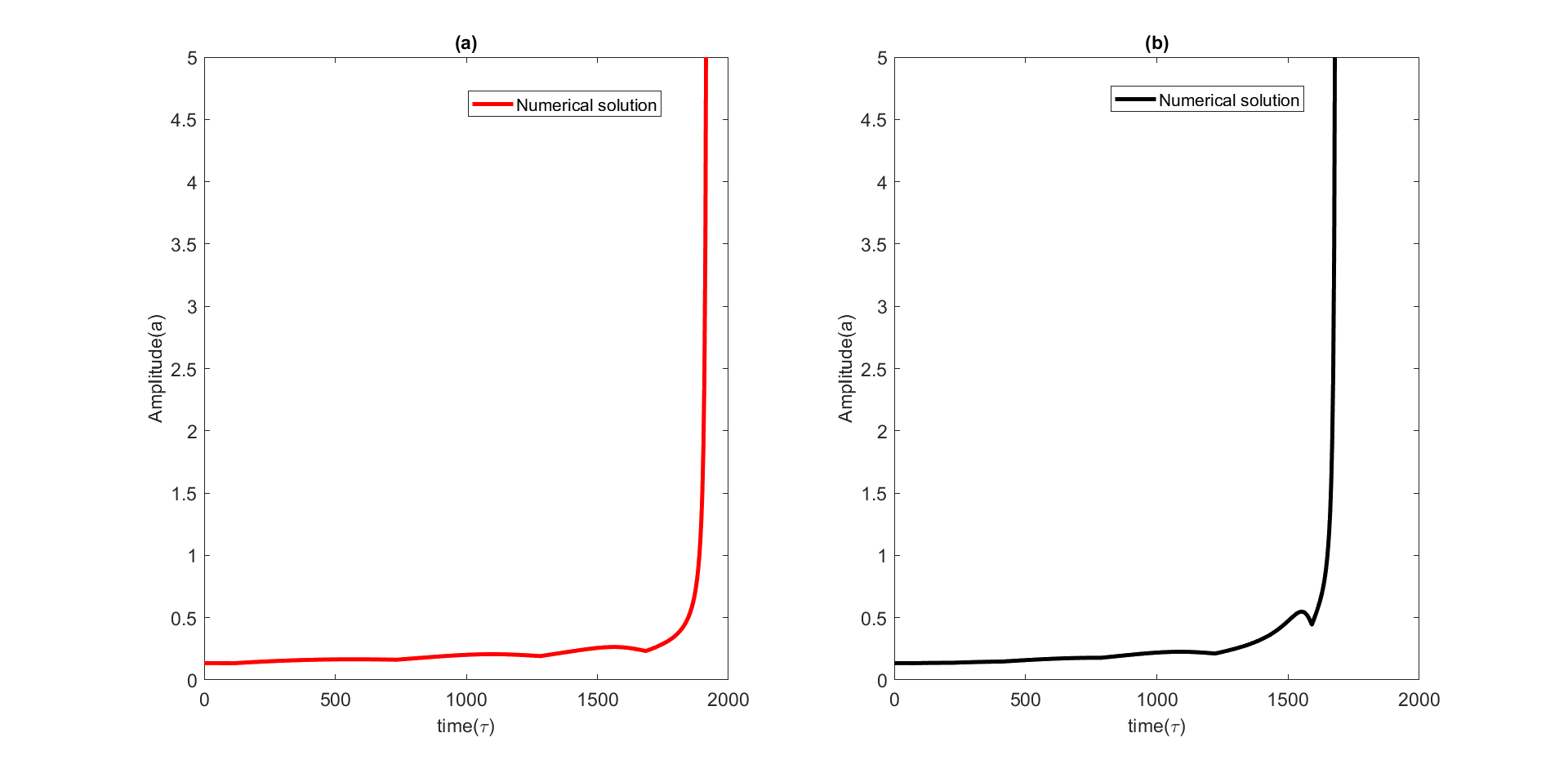}
	\caption{Examples of time evolution of the amplitude of a collapsing
		pulse with an initially Gaussian  distribution laterally
		stretched with the aspect ratio $\alpha=2$ (Solid red line)  and longitudinally stretched with $\alpha=\frac{1}{2}$ (Black solid line).
		Plot shows nondimensional amplitude $A$ against `slow time' $\tau, \,t=\varepsilon^{2}\tau$. Red solid line: simulated evolution of the
		amplitude of a collapsing
		pulse with the  initial
		condition ($a=0.1353,\sigma_{x}=25,
		\sigma_{y}=50$). Black solid line: simulated evolution of the
		amplitude of a collapsing
		pulse with the  initial
		condition ($a=0.1353,\sigma_{x}=50,
		\sigma_{y}=25$).
	}
\label{time_evolution_laterally-and-longitudinally stretched}
\end{figure}

A more nuanced complementary view of the evolution of the same pulse  is provided by the sequence of its cross-sections  in figure \ref{Cross-section_hairpin_vortices} and
a plot of the amplitude time dependence in figure (\ref{time_evolution_laterally-and-longitudinally stretched}). We note the following three key aspects of the evolution of collapsing pulses: \\(i) the amplitude growth accelerates with time, the amplitude becomes infinite at finite time, during the most of the pulse evolution the amplitude grows
very slowly, the sharp growth occurs just  prior to the
singularity;\\ (ii) the  center of the pulse moves forward accelerating with its velocity becoming infinite at the same moment as the amplitude becomes singular;\\
(iii) all pulses, i.e. with initially laterally or longitudinally stretched cross-sections,  become axially symmetric, although the 2D-BO evolution equation is not axially symmetric.

Qualitatively the
same pattern of evolution occurs for initial pulses with various shapes and aspect ratios. To illustrate the last
point,  we in figure \ref{time_evolution_laterally-and-longitudinally stretched} provide just a plot of the amplitude time dependence of a collapsing pulse for a longitudinally
stretched initial pulse with the aspect ratio $1/2$ .

Thus, we found that the localized perturbations blow up becoming axially symmetric in a vicinity of singularity. The simulations strongly suggest self-similarity of the evolution. In the next sections we derive and examine self-similar solutions of the 2D-BO equation and by comparing them with numerics will get a new insight into the evolution of the localized perturbations.

\section{Self-similar solutions of the 2D-BO equation}\label{Self-similarity}
	\subsection{Derivation of the general self-similar solution of the 2D-BO}\label{General-self-similarsol}
	If there is a singularity, it is natural to expect a self-similar behaviour of the
	solution in the vicinity of the singularity. In
	\cite{d1995two} a self-similar solution solution of the 2D-BO
	 was put forward. Here, we re-visit and modify their derivation.
	
Assuming that collapses  occur,  consider the spatial and temporal dynamics of a localized collapsing perturbation in a certain vicinity of the moving maximum of amplitude  presumed to be
at $\boldsymbol{r}_m(\tau)=\boldsymbol{r}_m\left\{x_m(\tau), 	y_m(\tau)\right\}$. We introduce  $\tau_{c}$,  the moment the singularity occurs and  $\check{\tau}=\tau_{c}-\tau$, the time  remaining to the
singularity.  In the frame of reference moving with  the maximum of amplitude
at $\boldsymbol{r}_m(\tau)$ we introduce new spatial coordinates $\check{\boldsymbol{r}}=
(\check{x}, \check{y})$, where 	$\check{x}=x-x_{m}(\check{\tau}),\, \check{y}=y-y_{m}(\check{\tau})$, that is, we make the moving position of the maximum the origin of the new frame $(\check{x}, \check{y})$.  Our
consideration differs from that of (\cite{d1995two}) by	taking into account time dependence of $\boldsymbol{r}_m(\check{\tau}) \{x_{m}(\check{\tau}), y_{m}(\check{\tau})\}$. This difference is crucial.
	
We look for self-similar solution in the form
	\begin{equation}\label{strat-self-similar-1}
		A(\check{\boldsymbol{r}},\check{\tau})=\check{\tau}^{\lambda_{1}}	h(\boldsymbol{\check{\xi}}),
		\quad
		\boldsymbol{\check{\xi}}=	\frac{\check{\boldsymbol{r}}}{\check{\tau}^{\lambda_{2}}},
		\quad \check{\boldsymbol{r}}=\check{x}\boldsymbol{i} +\check{y}\boldsymbol{j}=(x-x_{m}(\check{\tau}))\boldsymbol{i}+(y-y_{m}(\check{\tau}))\boldsymbol{j},
	\end{equation}
where $\boldsymbol{r}_m(\tau) $ remains to be specified. Note that the derivatives w.r.t  $\check{\tau}$ and  ${\tau}$ have opposite: $\dot x=\partial_{\check{\tau}}x$ and $\partial_{\tau}=-\partial_{\check{\tau}}$.

To transform each term in the evolution equation \eqref{IntroEvolutionEquation1} to the new spatial coordinate $\boldsymbol{\check{\xi}}$  we apply the chain rule to each term. To calculate the time derivative of ansatz \eqref{strat-self-similar-1},  we begin with the  term $\frac{\partial}{\partial\tau}h(\boldsymbol{\check{\xi}})=
	\frac{\partial h(\boldsymbol{\check{\xi})}}{\partial\boldsymbol{\check{\xi}}}
	\frac{\partial\boldsymbol{\check{\xi}}}{\partial\tau}$. Spelling out  $\frac{\partial\boldsymbol{\check{\xi}}}{\partial\tau}$ gives,
		\begin{equation}\label{eqnn-SLFsimilar2}
		\frac{\partial \boldsymbol{\check{\xi}}}{\partial\tau}=-\frac{\partial \boldsymbol{\check{\xi}}}{\partial\check{\tau}}=	 -\frac{\check{\tau}^{\lambda_{2}}(-\boldsymbol{\dot{\check r}}_{m})-\lambda_{2}\check{\tau}^{\lambda_{2}-1}\boldsymbol{r}}{\check{\tau}^{2\lambda_{2}}}=\check{\tau}^{-\lambda_{2}}
		\boldsymbol{\dot{r}}_{m}+\lambda_{2}\check{\tau}^{-1}\boldsymbol{\check{\xi}},\quad\text{where}\,\, \boldsymbol{\dot{r}}_{m}=\frac{\partial x_{m}}{\partial\check{\tau}}\boldsymbol{i}+\frac{\partial y_{m}}{\partial\check{\tau}}\boldsymbol{j},
	\end{equation}
Hence,  the  term $	\partial A/ \partial\tau$ in the evolution equation \eqref{IntroEvolutionEquation1} takes the form,
	\begin{equation}\label{eqnn-SLFsimilar-3}
		\frac{\partial A}{\partial \tau}=-\frac{\partial A}{\partial \check{\tau}}=-\frac{\partial}{\partial\check{\tau}}\check{\tau}^{\lambda_{1}}h(\boldsymbol{\check{\xi}})=\check{\tau}^{\lambda_{1}-1}[-\lambda_{1} h(\boldsymbol{\check \xi})+\lambda_{2}\boldsymbol{\check{\xi}} h^{\prime}(\boldsymbol{\check{\xi}})]+
		\check{\tau}^{\lambda_{1}-\lambda_{2}+\lambda_{3}}
		\boldsymbol{V}h^{\prime}(\boldsymbol{\check{\xi}}),
	\end{equation}
where $\boldsymbol{\dot{r}}_{m}\sim (V_{x}\boldsymbol{i}+V_{y}\boldsymbol{j})
\check{\tau}^{\lambda_{3}}=\boldsymbol{V}\check{\tau}^{\lambda_{3}}$. Here $\boldsymbol{V}$ is the notation we adopt for the time dependent velocity vector of the maximum of the pulse. Note that by taking time derivative of the ansatz   \eqref{strat-self-similar-1}  the time exponent $\lambda_{1}$ has been reduced
by one compared to the ansatz   in the first term and by $-\lambda_{2}+\lambda_{3}$ in the second term.
We  scale  the `velocities' $\dot{x}_{m}$ and $\dot{y}_{m}$ as $\check{\tau}^{\lambda_{3}}V_{m}^*$, where
$\lambda_{3}$ and $ V_{m}^*$ are unspecified yet  constants.  We presume the two terms of equation
\eqref{eqnn-SLFsimilar-3} to be dominant, i.e the terms of $O(\check{\tau}^{\lambda_{1}-1})$ to be of the same
order as terms of $O(\check{\tau}^{\lambda_{1}-\lambda_{2}+\lambda_{3}})$, and will verify this
assumption \emph{a posteriori}.

Next we find the nonlinear term $AA_{x}$  in  the new coordinates. Noting that $\frac{\partial}{\partial x}=\frac{\partial}{\partial\boldsymbol{\check{\xi}}}
	\frac{\partial\boldsymbol{\check{\xi}}}{\partial x}$, we find
		\begin{equation}\label{eqnn-NonlinearTerm}
		AA_{x}=\frac{1}{2}\frac{\partial}{\partial x}(A^{2})=\frac{1}{2}\frac{\partial }{\partial \boldsymbol{\check{\xi}}}(A^{2})\frac{\partial \boldsymbol{\check{\xi}}}{\partial x}=\frac{1}{2}\check{\tau}^{2\lambda_{1}-\lambda_{2}}\nabla\cdot(h^{2}(\boldsymbol{\check{\xi}}))\boldsymbol{i}=\check{\tau}^{2\lambda_{1}-\lambda_{2}}hh_{\boldsymbol{\check{\xi}}}\boldsymbol{i}\sim O(\check{\tau}^{2\lambda_{1}-\lambda_{2}}).
	\end{equation}
where $\nabla\equiv \frac{\partial}{\partial\boldsymbol{\check{\xi}}}$.
	
Similarly, we express the 2D-BO dispersion term  in new time and spatial coordinates to obtain:
	\begin{equation}\label{eqnn-Disp1}	 \hat{G}[A_{x}]=\check{\tau}^{\lambda_{1}-2\lambda_{2}}\hat{G}[h_{\boldsymbol{\check{\xi}}}]=\check{\tau}^{\lambda_{1}-2\lambda_{2}}\nabla\cdot\hat{G}[h(\boldsymbol{\check{\xi}})]\boldsymbol{i}\sim O(\check{\tau}^{\lambda_{1}-2\lambda_{2}}),
	\end{equation}
For the original evolution equation \eqref{IntroEvolutionEquation1} to remain invariant we assume that the two terms of equation \eqref{eqnn-SLFsimilar-3} are of the same order  and also that either of the two should balance the nonlinear term and the 2D-BO dispersion term given by  equation \eqref{eqnn-NonlinearTerm} and \eqref{eqnn-Disp1} respectively. This assumption of complete balance yields,
		\begin{equation}\label{eqnn-2DBO-selfsimilar-terms}
		\check{{\tau}}^{\lambda_{1}-1}[-\lambda_{1} h(\boldsymbol{\check\xi})+\boldsymbol{\check \xi}\lambda_{2} h^{\prime}(\boldsymbol{\check{\xi}})]+\check{\tau}^{\lambda_{1}-\lambda_{2}+\lambda_{3}}[\boldsymbol{V}^*_{m}h^{\prime}(\boldsymbol{\check{\xi}})]
 =\hat{G}[A_{x}]-AA_{x}=\check{{\tau}}^{\lambda_{1}-2\lambda_{2}}\hat{G}[h_{\boldsymbol{\check{\xi}}}]-\check{\tau}^{2\lambda_{1}-\lambda_{2}}hh_{\boldsymbol{\check{\xi}}}.
	\end{equation}
	To proceed,   we first balance the $\check\tau$ exponents  in the two terms in  $	\partial A/ \partial\tau$  given  by \eqref{eqnn-SLFsimilar-3} and then either of the two terms of $\partial A/ \partial \check\tau$ exponents with the exponents of the nonlinearity and  2D-BO dispersion. Upon rearrangement this ``distinguished limit'' leads to three equations which link  so far unspecified $\lambda_{1}$, $\lambda_{2}$ and $\lambda_{3}$, as follows
	\begin{subequations}\label{eqn-SimL1}
		\begin{equation}
			\lambda_{1}-1=\lambda_{1}-\lambda_{2}+\lambda_{3},\implies \lambda_{2}-\lambda_{3}=1,
		\end{equation}
		\begin{equation}
			\lambda_{1}-\lambda_{2}+\lambda_{3}=\lambda_{1}-2\lambda_{2},\implies \lambda_{2}+\lambda_{3}=0,
		\end{equation}	
		\begin{equation}
			\lambda_{1}-2\lambda_{2} =2\lambda_{1}-\lambda_{2},\implies  \lambda_{1}+\lambda_{2}=0.
		\end{equation}
	\end{subequations}
	The set of equations \eqref{eqn-SimL1} has unique  solution: $\lambda_{2}=\frac{1}{2},\,\lambda_{1}=\lambda_{3}=-\frac{1}{2}$ .
	On substituting these values of $\lambda_{i}$ into equation \eqref{strat-self-similar-1} we obtain the
explicit expression for the  self-similar solution obtained under the key assumption of the distinguished limit,
	\begin{equation}\label{B-eqnSLFsimilar1}
		A(\boldsymbol{r},\check{\tau})=\check{\tau}^{-\frac{1}{2}}	h(\boldsymbol{\check{\xi}}),
		\quad
		\boldsymbol{\check{\xi}}=	\frac{\check{\boldsymbol{r}}}{\check{\tau}^{\frac{1}{2}}},
	\end{equation}
On substituting the found values of $\lambda_{1},\lambda_{2}$ and $\lambda_{3}$ into equation \eqref{eqnn-2DBO-selfsimilar-terms} we obtain
	\begin{equation}\label{eqnn-2DBO-selfsimilar-terms-1} \frac{1}{2}[h(\boldsymbol{\check{\xi}})+\boldsymbol{\check{\xi}}h^{\prime}(\boldsymbol{\check{\xi}})]+
hh^{\prime}(\boldsymbol{\check{\xi}})\boldsymbol{i}+
\boldsymbol{V}_{m}^*h^{\prime}(\boldsymbol{\check{\xi}})-\hat{G}[h^{\prime}(\boldsymbol{\check{\xi}})]\boldsymbol{i}=0,
	\end{equation}
After some algebra we rearrange the terms of \eqref{eqnn-2DBO-selfsimilar-terms-1} and pull out the derivative with respect to $\boldsymbol{\check{\xi}}$ from the first two terms, while recalling that $h^{\prime}(\boldsymbol{\check{\xi}})\equiv \partial_{\boldsymbol{\check{\xi}}}h$ to obtain,
	\begin{equation}\label{eqnn-2DBO-selfsimilar-terms-2}		 \frac{1}{2}\partial_{\boldsymbol{\check{\xi}}}[\boldsymbol{\check{\xi}}h(\boldsymbol{\check{\xi}})+
h^{2}(\boldsymbol{\check{\xi}})\boldsymbol{i}]+\boldsymbol{V}_{m}^*h^{\prime}-
\hat{G}[h^{\prime}(\boldsymbol{\check{\xi}})]\boldsymbol{i}=0.
	\end{equation}
Upon integrating once the  resulting equation \eqref{eqnn-2DBO-selfsimilar-terms-2} we obtain a nonlinear equation on $h({\boldsymbol{\check{\xi}}})$ governing the spatial-temporal behaviour,
	\begin{equation}\label{A-spatial-dependence-2}
		(\boldsymbol{V}_{m}^*\boldsymbol{i})h+\frac{\boldsymbol{(\check{\xi}\cdot i})}{2}h+\frac{h^{2}}{2}-\hat{G}[h]=0.
	\end{equation}
Finally, upon further simplification we rewrite  the non-linear nonlocal equation \eqref{A-spatial-dependence-2} as
	\begin{equation}\label{anisotropic-selfsim-eqn}
		V_{m}^*h+\frac{\xi_{1}}{2}h+\frac{h^{2}}{2}-\hat{G}[h]=0.
	\end{equation}
	where $\xi_{1}=({x-x_{m})}/{\check{\tau}^{1/2}}$.
It is easy to see that our nonlocal nonlinear equation (\ref{anisotropic-selfsim-eqn}) specifying dependence on $h(\check{\boldsymbol{\xi}})$ is  anisotropic and, hence, does not support axially symmetric solutions, strongly suggested by our numerics discussed in the previous section. Thus, to describe axially symmetric patterns we need to look for  an alternative scaling.

	\subsection{Derivation of axially symmetric self-similar solutions}\label{axial-symmetry}

In the preceding section \S\ref{General-self-similarsol} we derived  a family of self-similar solutions assuming  all the temporal terms in equation \eqref{eqnn-SLFsimilar-3} to be in balance, which leads to a unique solution for $\lambda_{1},\lambda_{2}, \lambda_{3}$, that yields the self-similar solution governed by \eqref{anisotropic-selfsim-eqn}. This solution cannot be   axially symmetric and, hence contradicts our hypothesis that such a self-similar solutions can describe our numerical simulations in \S3.  Here, we consider a partial balance hypothesis: we assume  the term
 $\lambda_{3}\boldsymbol{V}_{m}^*h^{\prime}(\boldsymbol{\check{\xi}})$ in  equation \eqref{eqnn-SLFsimilar-3} to be dominant and  balance it with  both the nonlinearity and the dispersion terms. Thus, from equations \eqref{eqnn-SLFsimilar-3}, \eqref{eqnn-NonlinearTerm} and \eqref{eqnn-Disp1} we obtain a system of two linear equations with three unknown variables $\lambda_{1},\lambda_{2}$ and $\lambda_{3}$.
	\begin{subequations}\label{eqnn-linear-system1}
		\begin{equation}
			\lambda_{1}-\lambda_{2}+\lambda_{3}=2\lambda_{1}-\lambda_{2},\implies \lambda_{1}-\lambda_{3}=0,
		\end{equation}
		\begin{equation}
			\lambda_{1}-2\lambda_{2} =2\lambda_{1}-\lambda_{2},\implies  \lambda_{1}+\lambda_{2}=0.
		\end{equation}
		\end{subequations}
If we set $\lambda_{2}=\lambda>\frac{1}{2}$, where $\lambda$ is a free parameter,  then there are infinitely many solutions
to equations \eqref{eqnn-linear-system1} which admit axial symmetry. We re-iterate that in this regime the first
term of $O(\check{\tau}^{\lambda_{1}-1})$ in equation \eqref{eqnn-2DBO-selfsimilar-terms} is non dominant
and therefore drops out. Hence, upon integrating once we obtain the following nonlinear equation on
$h(\boldsymbol{\check{\xi}})$
	\begin{equation}\label{eqnn-spatial-symmetric}
		V_{m}^*h+\frac{h^{2}}{2}-\hat{G}[h]=0.
	\end{equation}
Equation  \eqref{eqnn-spatial-symmetric} first emerged as the stationary reduction of the 2D-BO equation in \cite{abramyan1992multidimensional}, where it was  examined numerically. The solutions for $h({\boldsymbol{\check{\xi}}})$ are specified by  equation \eqref{eqnn-spatial-symmetric} which is known to admit
axially symmetric solutions \cite{abramyan1992multidimensional}.

The resulting equation  \eqref{eqnn-spatial-symmetric}  can be  further simplified by assuming axial symmetry of the solution,  then the equation  can be reduced to a 	one-dimensional integral equation depending only on the radial 	coordinate. Indeed, the  2-D operator $\hat{G}[h]$ in the
presentation (\ref{Intro-eqGintegralOperatorCauchy})
retains its form in the $\check{x},\, \check{y}$ space,
	\begin{equation}\label{hkhat-integral-operator2}
		\hat{G}[h(\check{x},\check{y})]=\frac{1}{2\pi}\int_{-\infty}^{\infty}\int_{-\infty}^{\infty}
		\frac{h(\check{x}^{\prime},\check{y}^{\prime})\,d\check{x}^{\prime}\,
			d\check{y}^{\prime}}{[(\check{x}-\check{x}^{\prime})^{2}+
			(\check{y}-\check{y}^{\prime})^{2}]^{3/2}}.	
	\end{equation}
	
Assuming that $h$     depends on $\check{x},\, \check{y}$ only through radial  variable $ \check{r}=\sqrt{\check{x}^2+\check{y}^2}$, we can transform \eqref{hkhat-integral-operator2} into polar coordinates and then to integrate it with respect to the polar angle. Then  the 2-D  equation \eqref{eqnn-spatial-symmetric} specifying $h(\boldsymbol{\check{\xi}})$    reduces to  the novel one-dimensional nonlinear Fredholm integral equation of the second kind depending only on the self-similar radial coordinate, 	\begin{equation}\label{radial}
	V_{m}^*h+\frac{h^{2}}{2}-\hat{G_1}[h] =0, \quad 	\hat{G}_{1}[h(\check \xi,\tau)]=\frac{2}{\pi}	 \int_0^{\infty}\frac{h({\check \xi}^{\prime},\tau)\, {\check \xi}^{\prime}\,E(\gamma^{\prime})\,d\check \xi^{\prime}}{(\check \xi^{\prime}-\check \xi)^{2}({\check \xi}+{\check \xi}^{\prime})},
	\end{equation}
where $E(\gamma^{\prime})$ is the incomplete elliptic integral of the second kind (e.g. Olver et al 2010). This is the exact reduction of \eqref{eqnn-spatial-symmetric}. We introduced operator $\hat{G_1}[h]  $ to designate  the axially symmetric reduction of $\hat{G}[h] $. The details of this reduction are given in Appendix  \ref{G1-operator-transformation}.

\begin{figure}
		\centering
		\includegraphics[width=1.0\textwidth]
		{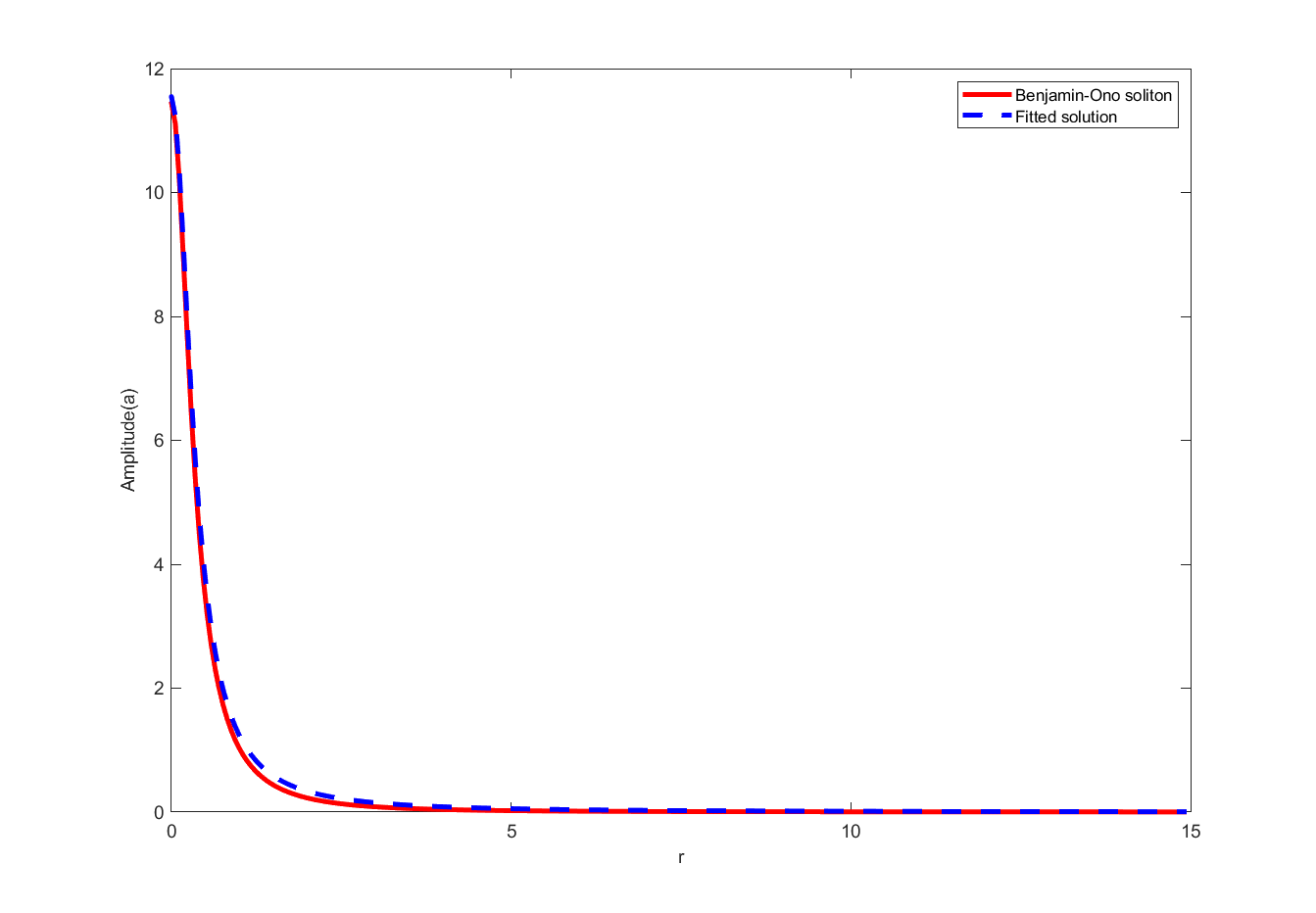}
		\caption{Exact numerical
			solution of \eqref{eqnn-spatial-symmetric} (blue dashed line) and an example of the   `radial' BO
soliton approximation in $\xi$ variable (solid red line) and the superimposed upon it. \%It shows  how well the Lorentzian pulse BO solution  captures
		}
		\label{Benjamin-ono-soliton}
	\end{figure}

At the moment we cannot solve analytically the resulting  hypersingular 1-D integral equation \eqref{radial}. However, we examine it numerically. The equation has only one free parameter $ V_{m}^*$. Remarkably, a straightforward  `radialization' of  Benjamin-Ono soliton proved to provide a quite good fit to the numerical  solution
	\begin{equation}\label{BO-soliton}
		h(\check{\boldsymbol \xi})\approx \frac{4V_{m}^*}{1+ {V_{m}^*}^2{\check{|\xi |}^2}},  \end{equation}
This axially symmetric 1-D Benjamin-Ono soliton has been obtained just by extrapolating 1-D Benjamin-Ono soliton formula to the axially symmetric context, it is a guess, not a rational approximation.
Figure (\ref{Benjamin-ono-soliton})  shows how well the BO soliton  \eqref{BO-soliton} captures  the exact
numerical solution of  equation  \eqref{radial}.  The single parameter of the solution, the arbitrary constant
$V_{m}^*$  characterizes both the width and amplitude of the pulse, for more details see Appendix
 \ref{G1-operator-transformation}.

In the variables $\check{\boldsymbol{r}},\check{\tau}$ the
 radial Benjamin-Ono  approximation reads
 \begin{equation}\label{BO-soliton-in-r--tau}
		h(\check{\boldsymbol{r}},\check{\tau})\approx
\frac{4V_{m}^*}{1+ {V_{m}^*}^2\frac{|\boldsymbol{\check{r}}|^2}
{\check{\tau}^{2\lambda}}},  \qquad A\approx \check{\tau}^{-\lambda}
\frac{4V_{m}^*}{1+ {V_{m}^*}^2\frac{|\boldsymbol{\check{r}}|^2}
{\check{\tau}^{2\lambda}}}.
\end{equation}
This representation clarifies the place of the radial Benjamin-Ono  soliton-like solutions \eqref{BO-soliton-in-r--tau} in the big picture: they are analogues of the so-called Townes' envelope solitons in 2-D systems \cite{chiao1964self}. The radial Benjamin-Ono  soliton-like solutions \eqref{BO-soliton-in-r--tau} describe self-similar collapsing pulses similar to the  Townes' envelope solitons describing 2-D collapses.

We summarize the results of this section as follows: we found two families of self-similar solutions of
the 2D-BO equation: the first one is not axially symmetric, it is obtained under
assumption of complete balance which  uniquely determines the  exponents $\lambda_i$ in the assumed self-similar
ansatz; the second family of solutions   \eqref{strat-self-similar-1} is axially symmetric,   it is obtained
under assumption of partial balance which  leaves  undetermined exponent $\lambda_i$ as a free parameter
in the solution. We cannot uniquely determine the exponent within this approach. The physical sense  of the
found solutions and their place in reality is similar to that of the  Townes envelope solitons.

\section{Self-similar solutions vs numerical simulation of initial problem}\label{Quantative Comparisons}
The simulations of \S\ref{Collapses in 2dBO} strongly suggest that localized perturbations collapse in a self-similar way apparently becoming axially symmetric. The analysis of self-similar solutions in \S\ref{Self-similarity}  showed that there are two very distinct families of self-similar solutions. Here, by re-examining the initial  problem for an initial pulse we are trying to address the following basic questions: \begin{description}
                                                                        \item[(i)] Can any of the found self-similar solutions be observed in the simulations? If yes, what predictions of the self-similar solutions are confirmed by the simulations.
                                                                        \item[(ii)] Can the exponents be quantified? How do parameters of the initial perturbations manifest in the self-similar stage of the solution?
                                                                      \end{description}

\begin{figure}[h!]
	\centering
	\includegraphics[width=1.0\textwidth]
	{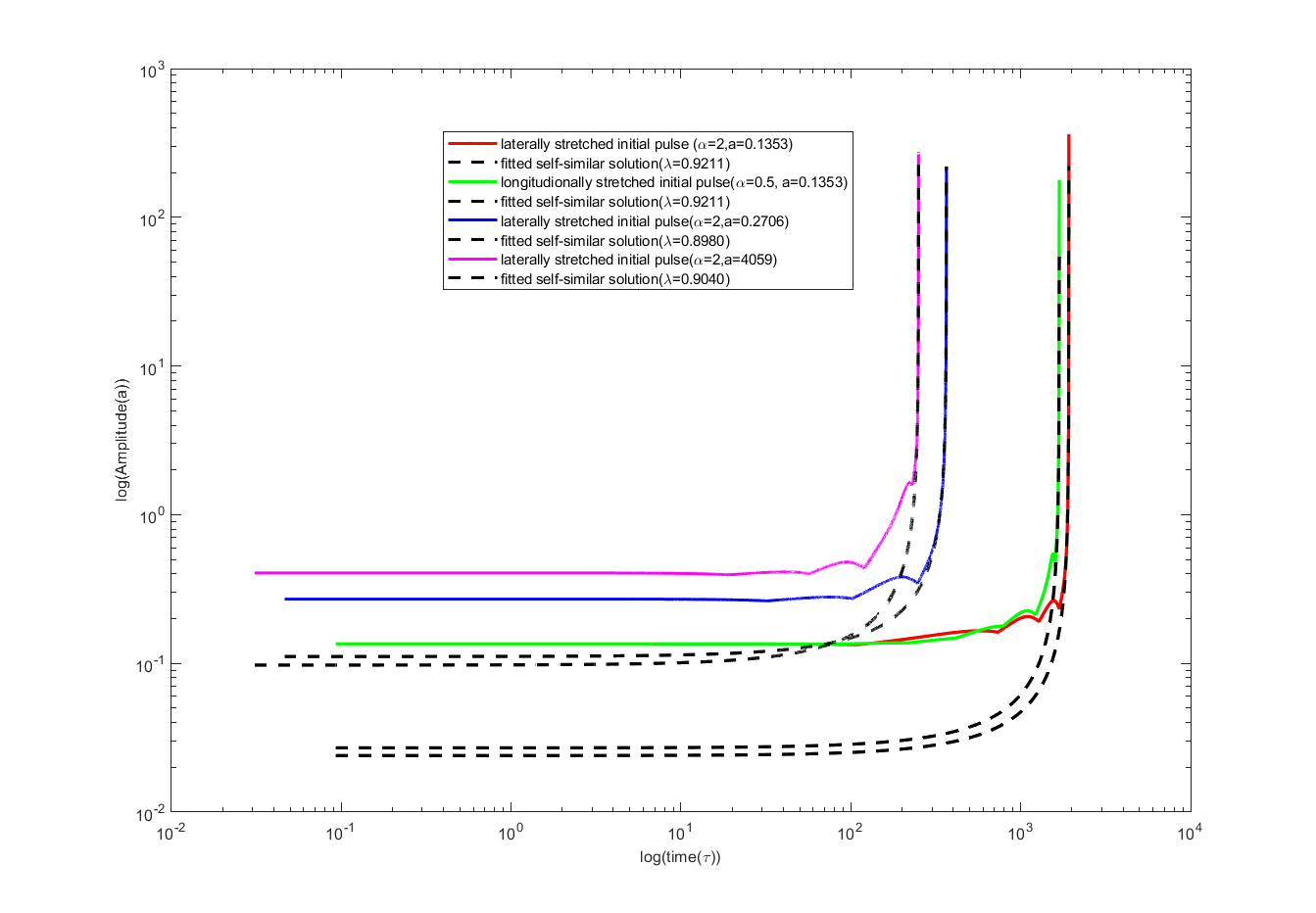}
	\caption{Loglog scale plots of amplitude evolution of  collapsing
		pulses of various initial asymmetry and amplitude: solid lines --simulated evolution,
dashed  lines --fitted self-similar solutions.\\
		Red solid line: evolution of the
		amplitude of a collapsing initially laterally stretched Gaussian
		pulse. The  parameters of the initial distribution prescribed (\ref{B-InitialConditions}): $a=0.1353,\sigma_{x}=50,
		\sigma_{y}=25$.\\
		Green solid line: simulated evolution of the
		amplitude of a collapsing initially longitudinally stretched  Gaussian
		pulse (\ref{B-InitialConditions}) with the  parameters
		($a=0.1353,\sigma_{x}=25,
		\sigma_{y}=50$).
\\Blue solid line (twice increased  amplitude): $a=0.2706, \sigma_{x}=50,
		\sigma_{y}=25$. \\Magenta solid line (thrice increased amplitude: $a=0.4059, \sigma_{x}=50,
		\sigma_{y}=25$)\\
		Dashed lines: their matched self-similar solutions.
	} \label{amplitude-time-loglog-plot}
\end{figure}

Figure (\ref{amplitude-time-loglog-plot}) shows loglog plots of amplitude vs time in   several cases of collapses  simulated numerically and compared to their  counterparts obtained by fitting the axially symmetric self-similar solutions.  Exponents $\lambda$ are found by fitting the simulated curve. The figure is plotted for different values of initial amplitude $a$ and asymmetry parameter $\alpha$; the axially symmetric self-similar solution with time exponent parameter $\lambda$ is fitted independently for each case.  It is easy to see that the self-similar solution captures very well the amplitude evolution as the perturbations approach collapse. It is also evident that the exponent of the self-similar solution does not depend on the asymmetry parameter $\alpha$ nor amplitude $a$ in the initial condition. The discrepancy in the found values of  $\lambda$ is $(10^{-2})$ for different initial amplitudes and asymmetry parameter $\alpha$. We stress that found exponent  $\lambda \approx 0.9$ is far from the value  $\lambda=1/2$ predicted by the self-similar solutions obtained under assumption of complete balance. To better illustrate the scatter in the values of $\lambda$ we provide a table (see \ref{table-1}) quantifying the error margin. The simulations are performed with 95\% confidence interval, hence the error margin is $\pm 0.05$. A summary of the initial values of amplitudes, asymmetry parameter and corresponding $\lambda$ are summarized in table \ref{table-1}.\\

\begin{table}[h!]
	\centering
	\caption{Self-similar solution time exponent  $\lambda$ found by fitting and the margin of error for different   initial conditions}
	\label{table-1}
	\begin{tabularx}{0.8\textwidth} {
		| >{\raggedright\arraybackslash}X
		| >{\centering\arraybackslash}X
		| >{\raggedleft\arraybackslash}X| }
	\hline
	Initial widths $(\sigma_{x},\,\sigma_{y})$ and the aspect ratio $\alpha$& Initial amplitude($a$) & $\lambda$ and the error margin  \\
	\hline
	$\alpha=\frac{1}{2}$  & $0.1353$  & $0.9211\pm 0.05$  \\
	\hline
	$\alpha=2$  & $0.1353$& $0.9211\pm 0.05$ \\
	\hline
	$\alpha=2$  & 0.2706  & $0.8980\pm 0.05$ \\
	\hline
	$\alpha=2$  & 0.4059  & $0.9040\pm 0.05$ \\
	\hline
\end{tabularx}
\end{table}

 \begin{figure}[h!]
		\centering
		\includegraphics[width=1.0\textwidth]
		{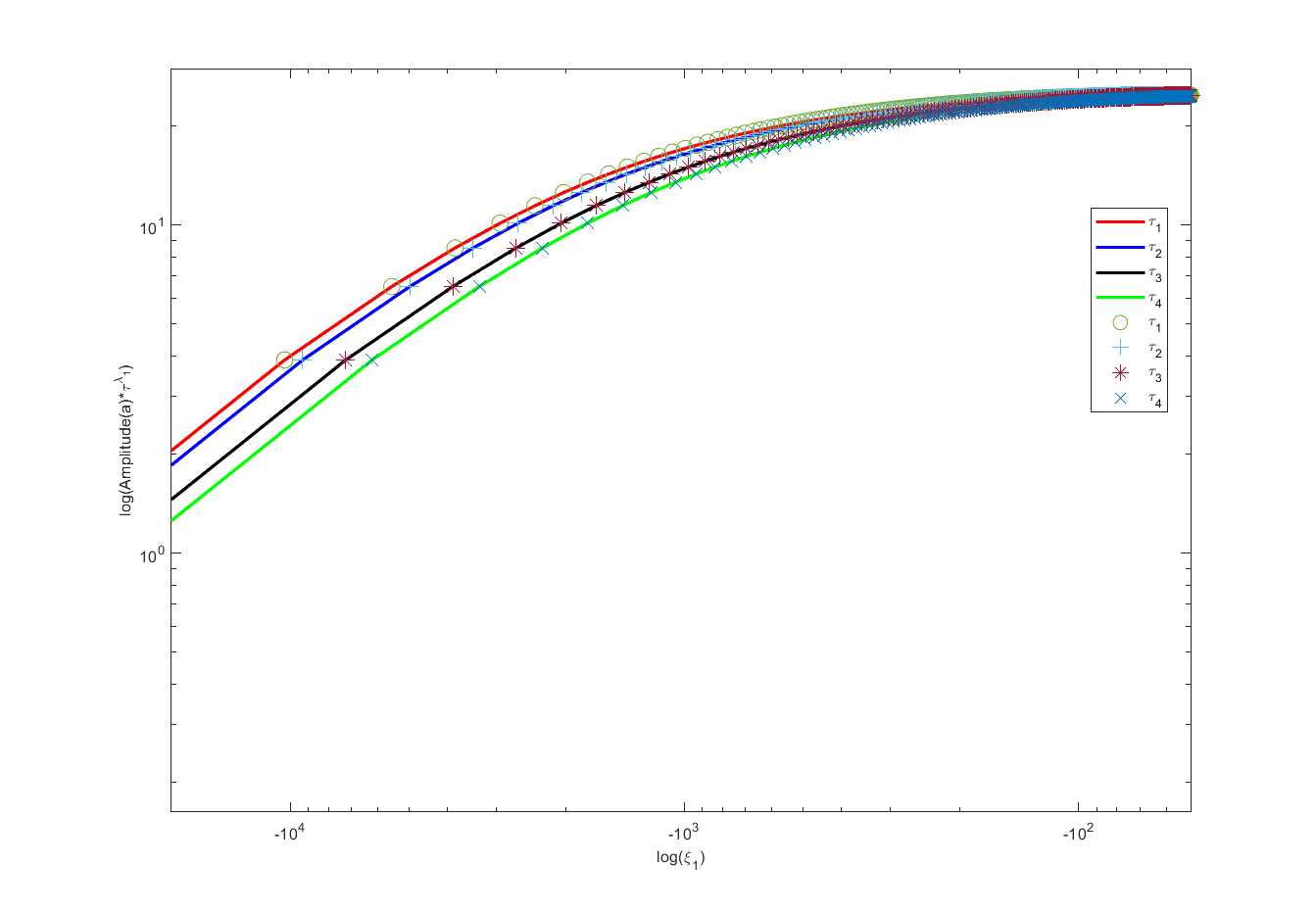}
		\caption{Scaled loglog  plots of amplitude evolution of  collapsing
				pulses in the self-similar variables for different  moments: ($\tau_{1}=92,\,\tau_{2}=94,\,\tau_{3}=96,\,\tau_{4}=98$). Solid lines dependence on the scaled longitudinal variable $\log \xi_1$, dependence on the transverse scaled variable $\log \xi_2$ is shown by superimposed symbols. 
			} \label{selfsim-loglog-plot}
	\end{figure}
Thus, figures (\ref{amplitude-time-loglog-plot}, \ref{selfsim-loglog-plot}) show that near the singularity evolution of collapsing perturbations is described by the axially symmetric self-similar solutions. Figure ( \ref{selfsim-loglog-plot})  confirms that the evolution of collapsing pulses indeed becomes axially symmetric, since the longitudinal and transverse dependencies coincide. The exponent  $\lambda$  found by fitting proved to be remarkably robust (approximately equal to 0.9), and, as we  could see from the examples we simulated, independent of the asymmetry and amplitude of the initial pulse.

	\section{Concluding Remarks}\label{Main-Conclusions}
	Our main results can be briefly summarized as follows. By simulating numerically the 2D-BO equation  we found 	that  the collapsing localized initial perturbations, i.e. those  with negative values of the Hamiltonian, become axially symmetric in the course of evolution, whatever their initial asymmetry. The simulations also strongly suggest  self-similar character of the evolution. The 2D-BO equation does not posses rotational  symmetry,  its family of self-similar solutions obtained under the assumption of complete balance does not admit axially symmetric solutions either. To explain the results of the simulations we assumed partial balance and under this assumption found a family of axially symmetric self-similar solutions. \emph{A priori } assuming axial symmetry of the solution, we obtained a novel integral equation describing collapsing solitary-type patterns which are similar in their role to the `Townes' solitons'.  This  novel integral equation might be viewed as a one-dimensional model of collapse. Mathematically, the steady soliton-like solutions in terms of self-similar variables describing the axially symmetric collapse are the axially symmetric solitons found in \cite{abramyan1992multidimensional} as steady solutions of  the 2D-BO equation. Our analysis of the evolution of various initially localized  pulses shows that in a certain vicinity of the singularity the pulses are behaving in accordance with the predictions of our axially symmetric self-similar solution,  the unspecified  exponent  $\lambda$ in the self-similar solution is found by fitting the analytical and numerical solitons. The exponent   found by fitting,  $\lambda \approx 0.9$, proved to be remarkably robust  and independent of the asymmetry and amplitude of the initial pulse, at least for the range of the examples we simulated.

The results above raise a  number of questions.   Currently we do not know
why  the axially symmetric self-similar solutions emerge out of arbitrary initial conditions despite the lack of axial symmetry in   the 2D-BO equation.  How is a particular exponent selected? What is the mechanism? Why is exponent  $\lambda$ almost universal? Apparently it does not depend, or depends very weakly, on the initial conditions. How are the parameters of the axially symmetric self-similar solutions related to the initial conditions? There are also fundamental questions lying beyond the  2D-BO equation framework. The 2D-BO equation  is a weakly nonlinear asymptotic model, we do not know the eventual outcome of the found
collapses in the full Navier-Stokes equations. We are currently working on  clarifying these outstanding
questions, resolving them will considerably advance our understanding of  physics of collapses.\\

	\textbf{Acknowledgements}\\
	The authors are grateful to Prof. C.J.Chapman for helpful comments and discussions.
\\
	
	\textbf{Conflicts of Interest}\\
	The authors declare that there is no conflict of interest.
	\appendix
	\section{Analysis of the 2D-BO integral operator $\hat{G}[h]$ for axisymmetric pulses}\label{G1-operator-transformation}
	The Appendix details  axisymmetric reduction of the 2D-BO integral operator.
	In \S\ref{Axisymmetric reduction} we first perform  transformation of the 2D-BO integral operator \eqref{hkhat-integral-operator2} from the Cartesian frame into polar coordinates,  which reduces  the 2D-BO to a one dimensional hypersingular integral equation. The hypersingular integral is understood in the Cauchy-Hadamard sense.
	In \S\ref{Inner} we further simplify the integrand.The results are briefly summarized in  \S\ref{A-concl}.
	\subsection{Axisymmetric reduction of the 2D-BO integral operator $\hat{G}[h]$}\label{Axisymmetric reduction}
	
	We begin with the general form of the stationary 2D-BO equation first derived and examined in 	 \cite{abramyan1992multidimensional}. In \S\ref{Self-similarity},  \ref{Quantative Comparisons} it was shown  that the 	evolution of collapses is self-similar and that the instantaneous  spatial distribution described by the self-similar
	solution is governed by the steady 2D-BO  equation 
	\begin{equation}\label{stationary 2-D BO}
		V_{m}^{*}h+\frac{h^{2}}{2}-\hat{G}[h]=0,
	\end{equation}
	where $V_{m}^{*}$ is a constant, while the independent Cartesian variables $\xi_{1}, \, \xi_{2}$ are  $\xi_{1}=\frac{(x-x_{m}(\tau))}{\tau^{\lambda}}$ and $\xi_{2}=\frac{(y-y_{m}(\tau))}{\tau^{\lambda}}$. Consider the dispersion integral operator employing traditional $x,y$ notation:  $\hat{G}[h(x,y)]$. The dispersion integral operator $\hat{G}[h(x,y)]$  is the hypersingular integral understood as the Cauchy-Hadamard integral
		\begin{equation}\label{Ghat-operator-1}
			\hat{G}[h(x,y)]=\frac{1}{2\pi}\int_{-\infty}^{\infty} \int_{-\infty}^{\infty}	 \frac{h(x^{\prime},y^{\prime})\,dx^{\prime}\,dy^{\prime}}{[(x-x^{\prime})^{2}+(y-y^{\prime})^{2}]^{3/2}}.
		\end{equation}
		The  numerical considerations of  \S\ref{axial-symmetry} strongly suggest a robust  tendency of localized solutions towards axial symmetry. It is therefore tempting to assume \emph{a priori } axial symmetry and exploit the simplification provided by the symmetry. To proceed, we first transform the integral operator \eqref{Ghat-operator-1} given in in terms of Cartesian coordinates $x,y,x^{\prime}, y^{\prime}$ into  the polar coordinates $r, \theta$
		$$x=r\cos{\theta},\quad y=r\sin{\theta},\quad x^{\prime}=r^{\prime}\cos{\theta^{\prime}},\quad y^{\prime}=r^{\prime}\sin{\theta^{\prime}}.$$
		Then we express the binomials under the square root  in polar coordinates
		\begin{equation}\label{polar-x}
			(x-x^{\prime})^{2}=(r\cos{\theta}-r^{\prime}\cos{\theta^{\prime}})^{2}=
			 r^{2}\cos^{2}{\theta}-2rr^{\prime}\cos{\theta}\cos{\theta^{\prime}}+r^{\prime\,2}\cos^{2}{\theta^{\prime}},
		\end{equation}
		
		\begin{equation}\label{polar-y}
			(y-y^{\prime})^{2}=(r\sin{\theta}-r^{\prime}\sin{\theta^{\prime}})^{2}=
			r^{2}\sin^{2}{\theta}-2rr^{\prime}\sin{\theta}\sin{\theta^{\prime}}+r^{\prime\,2}
			\sin^{2}{\theta^{\prime}},
		\end{equation}
		and  sum-up equations \eqref{polar-x} and \eqref{polar-y}. Using standard trigonometric identities we obtain
		\begin{equation}\label{polar-x-y}
			(x-x^{\prime})^{2}+(y-y^{\prime})^{2}=r^{2}+r^{\prime\,2}-
			2rr^{\prime}\cos{(\theta^{\prime}-\theta)}.
		\end{equation}
		On substituting equation \eqref{polar-x-y} into \eqref{Ghat-operator-1}  and noting that $dx^{\prime}dy^{\prime}=r^{\prime}dr^{\prime}d\theta^{\prime},$
		we find
		\begin{equation}\label{khat-integral-operator1}
			\hat{G}[h(\boldsymbol r)]=\frac{1}{2\pi}\int_0
			^{\infty}dr^{\prime}  \int_{0}^{2\pi} \frac{h(r^{\prime}\cos{\theta^{\prime}},r^{\prime}\sin{\theta^{\prime}},\tau)\,r^{\prime}d\theta^{\prime}}{[r^{2}+r^{\prime\,2}-
				2rr^{\prime}\cos{(\theta^{\prime}-\theta)}]^{3/2}}.
		\end{equation}
		To distinguish usage of operator $\hat{G}[h(\boldsymbol r)]$ in the full 2D space $x,y$ from the situations where it is applied to the class of axially symmetric functions we introduce a new notation  $ \hat{G}_{1}[h(r)] $ for the latter.
		
		\subsection{Inner integral $I_{1}$}\label{Inner}
		First, we re-write and investigate the inner integral which we denote here as $I_1$. The integration   is with respect to the dummy variable $\theta^{\prime}$. Without loss of generality we set $\theta=0$. Finally, we present  the inner  integral in  the following compact form,
		\begin{equation}\label{Integral-I}
			I_1=\int_{0}^{2\pi}\frac{d\theta^{\prime}}{\sqrt{(a-b\cos{\theta^{\prime}})^{3}}},\qquad a=r^{2}+r^{\prime\,2},\quad b=2rr^{\prime}.
		\end{equation}
		Consider the inner integral $I_{1}$ given by equation \eqref{Integral-I} after renaming the dummy variable $\theta^{\prime}$ by $x$ to obtain,
		\begin{equation}\label{new-integral-I1} I_{1}=\int_{0}^{2\pi}\frac{dx}{(m-n\cos{(x)})^{\frac{3}{2}}}=2\int_{0}^{\pi}\frac{dx}{\sqrt{(m-n\cos(x))^{3}}},\qquad m=r^{2}+r^{\prime\,2},\quad n=2rr^{\prime}.
		\end{equation}
		The elliptic integral \eqref{new-integral-I1}  can be expressed as (see e.g. \cite{zwillinger2007table} p. 182 equation 291.01) ,
		\begin{equation}
			I_{1}=\frac{2}{(m-n)(\sqrt{(m+n)})}E(\delta,\gamma^{\prime}),
		\end{equation}
		where   $E(\delta, \gamma^{\prime} )$ is the incomplete elliptic integral of the second kind. The parameters $\delta$ and the modulus of elliptic integral $\gamma^{\prime}$ are
		$$\delta=\sin^{-1}\left(\sqrt{(\frac{(m+n)(1-\cos{x})}{2(m-n\cos{x})})}\right),\quad \gamma^{\prime}=\sqrt{\frac{2n}{m+n}},\quad m>n>0,\quad 0\leq x\leq \pi.$$
		On substituting the above expressions for $m$ and $n$ into equation \eqref{new-integral-I1},  while making use of the limits of integration and  the property  of incomplete elliptic integral of the second kind, that $E(\frac{\pi}{2},\gamma^{\prime})=E(\gamma^{\prime})$ (see e.g. \cite{olver2010nist} p. 491 equation 19.6.9), we  simplify the expression for $I_{1}$,
		\begin{equation}
			I_{1}=\frac{4}{(r^{\prime}-r)^{2}(r+r^{\prime})}E(\gamma^{\prime}),\qquad \gamma^{\prime}=\frac{2\sqrt{rr^{\prime}}}{r+r^{\prime}},
		\end{equation}
		where $E(\gamma^{\prime})$ is the complete elliptic integral of the second kind. On substituting this result  for $I_{1}$ into equation \eqref{khat-integral-operator1} we express the 2D integral operator in  one-dimensional form in terms of $r$ as,
		\begin{equation}\label{khat-integral-operator-Radial-2}
			\hat{G}_{1}[h(r)]=\frac{2}{\pi}
			\int_0
			 ^{\infty}\frac{h(r^{\prime},\tau)\,r^{\prime}\,E(\gamma^{\prime})\,dr^{\prime}}{(r^{\prime}-r)^{2}(r+r^{\prime})}.
		\end{equation}
		This is the exact axisymmetric  reduction of integral \eqref{Ghat-operator-1}.	The properties of the complete elliptic integral of the second kind $E(\gamma^{\prime})$ are well known, in particular, we note that $E(1)=1$ and $E(0)=\frac{\pi}{2}$ (Olver et. al 2010).
		
		Thus, the problem of solving 2D-BO stationary equation \eqref{stationary 2-D BO} in the class of axisymmetric functions has been reduced to solving one-dimensional nonlinear   Fredholm integral equation of the second kind
		\begin{equation}\label{radial Fredholm}
			V_{m}^{*}h+\frac{h^{2}}{2}-\hat{G}_{1}[h]=0,
		\end{equation}
		where $\hat{G}_{1}[h]$ is given by \eqref{khat-integral-operator-Radial-2}. By construction the numerical solution of \ref{radial Fredholm} coincides with that of \ref{eqnn-spatial-symmetric}.
		
		Figure  (\ref{Benjamin-ono-soliton})  shows the soliton-like numerical solution of \eqref{stationary 2-D BO} obtained in \cite{abramyan1992multidimensional} and its Benjamin-Ono fit. It illustrates how well the Benjamin-Ono fit   \eqref{BO-soliton} given by the expression
		\begin{equation}\label{radial-spatial-solution}
			h(r)=\frac{4a_{0}}{1+a_{0}^{2}r^{2}},
		\end{equation}
		captures  the exact numerical solution of  equation  \eqref{stationary 2-D BO}  or its equivalent \eqref{radial}.
		Here $a_{0}$ is a constant characterizing both the amplitude and width. It is obtained numerically by fitting
		the cross-section and given by the parameter  $a_{0}=V_{m}^{*}=2.8876$ with over 95\% confidence interval.
		This is an illustration of the so-called the `ground  mode' solution,  there are also solitary wave solutions with oscillatory tails which we do not consider here.
		
		\subsection{Conclusions }	\label{A-concl}
		We have simplified the integral equation governing the spatial structure of collapses and found that  the radial Benjamin-Ono soliton fit captures  well the exact numerical solution. However, the problem of finding the solution analytically with a desired accuracy remains open.
	
	\bibliography{physicaDSCT}
	
	\end{document}